\documentclass[superscriptaddress, showpacs,showkeys,twocolumn,showpacs,amsmath,amssymb]{revtex4}
\usepackage{graphicx}
\usepackage{dcolumn}
\usepackage{bm}
\begin{document}
\title{
Power flow tracing in a simplified highly renewable European electricity network
}

\author{Bo Tranberg}
\affiliation{%
Department of Physics, Aarhus University, Ny Munkegade 120, 8000 Aarhus C, Denmark
}
\author{Anders B. Thomsen}
\author{Rolando A. Rodriguez}
\affiliation{%
Department of Mathematics, Aarhus University, Ny Munkegade 118, 8000 Aarhus C, Denmark
}
\author{Gorm B. Andresen}
\affiliation{%
Department of Engineering, Aarhus University, Inge Lehmanns Gade 10, 8000 Aarhus C, Denmark
}
\author{Mirko Sch\"{a}fer}
\affiliation{%
Frankfurt Institute for Advanced Studies, Ruth-Moufang-Str.\ 1, 60438 Frankfurt, Germany
}
\author{Martin Greiner}
\affiliation{%
Department of Engineering, Aarhus University, Inge Lehmanns Gade 10, 8000 Aarhus C, Denmark
}

\date{\today}
\begin{abstract}
The increasing transmission capacity needs in a future energy system raise the question how associated costs should be allocated to the users of a strengthened power grid. In contrast to straightforward oversimplified methods, a flow tracing based approach provides a fair and consistent nodal usage and thus cost assignment of transmission investments. This technique follows the power flow through the network and assigns the link capacity usage to the respective sources or sinks using a diffusion-like process, thus taking into account the underlying network structure and injection pattern. As a showcase, we apply power flow tracing to a simplified model of the European electricity grid with a high share of renewable wind and solar power generation, based on long-term weather and load data with an hourly temporal resolution.
\end{abstract}


\maketitle
%
%
\section{Introduction}
\label{sec:1}
The main drivers of the increasing integration of renewable energy into present and future energy systems are climate change mitigation and sustainability~\cite{Sims11}. Due to its fluctuating and decentralized nature, the renewable energies represent a challenge in terms of security of supply, robustness and cost efficiency. In this regard, detailed techno-economic models of the energy system are invaluable for the evaluation of well-defined scenarios, but often lack the possibility to assess fundamental mechanisms and interrelations due to their sheer complexity and multitude of parameters. On the other hand, oversimplified abstract models often neglect fundamental correlations and structures present in real systems. At this point, the applied theoretical physics of complex renewable energy networks represents an important bridge between these poles~\cite{hei10, hei11, ras12, jen14, rod15}. The incorporation of suitably aggregated long-term weather-determined renewable generation and historic load data accounts for the relevant spatio-temporal correlations, while the usage of simplified network and dispatch models reduce the complexity of the problem and limits the number of parameters. Such an approach allows to identify fundamental interrelations, develop and test new concepts -- often inspired from physics -- and thus is able to guide the research of more in-depth models. This point of view is shared by several recent contributions from the physics community addressing challenges for modern complex electricity infrastructures, with a strong focus on power grid synchronization, stability, and control~\cite{fila08, nishi15, rohd12, mott13, schmiet14, menck14, whaut12, schaef15, manik14}.

Following this line of research, in~\cite{TYNDP, rod14, bec14} it has been shown that for a highly renewable European electricity network, the overall transmission capacity needs can become quite large. In this respect, a simple, but important question emerges: who is going to pay for this? Straightforward propositions distribute the overall transmission capacity onto the nodes proportional to their average load, or count only the links $l(n)$ attached to a node~$n$. Although the average load is a first approximation of the respective network usage, it does not take into account the geographical position of a node in the network, nor any correlations in the import/export distribution of the system. Power flow tracing incorporates such relationships and thus is a vital ingredient of a more appropriate usage measure underlying a ``fair'' and consistent cost allocation procedure. 

This technique, also called average participation, has been pioneered by~\cite{bia96} and~\cite{kir97, str98}. It has been developed and suggested for transmission congestion management and pricing~\cite{sha02}, for instance in the context of inter-Transmission System Operator (TSO) compensations, that is the tariffication of cross-border flows between different European grid zones~\cite{con06}. Applying the principle of proportional sharing, the tracing algorithm acts on a given power flow solution and provides a partition of the total flow on the network into different partial flow patterns emerging from the respective sources, or alternatively ending in the respective sinks. These partial flow patterns can be represented by colours according to their originating (alternatively ending) node. Technically, flow tracing then corresponds to colour diffusion on a directed flow graph embedded in the original network.

In this paper we introduce a new link usage measure based on the flow tracing method and apply it to a simplified model of a highly renewable European electricity network. Using a notation based on propagating colour vectors, we reformulate the tracing technique using concepts familiar from diffusion processes on complex networks~\cite{new10}. Considering an ensemble of different flow patterns on a given network, the flow tracing techniques yield an ensemble of coloured partial flow patterns. The new usage measure integrates this information into a single value for each source, sink and link in the network. The application of these methods to the European electricity network then provides information about the respective grid usage of different countries, and the composition of their power imports and exports. Using appropriate cost functions, this usage measure can be integrated into future cost allocation schemes which provide a ``fair'' pricing of network related costs to individual importers and exporters.

This paper is organized in the following manner: Section~\ref{sec:2} provides a simplified description of renewable energy networks. Flow tracing on a network is the topic of Section~\ref{sec:3}. Section~\ref{sec:4} applies flow tracing to a simplified highly renewable European electricity network, and includes an ensemble averaging over fluctuating flow configurations. A conclusion and outlook is given in Section~\ref{sec:5}.

%
\section{Renewable electricity networks}
\label{sec:2}

\begin{figure}
\includegraphics[width=0.5\textwidth]{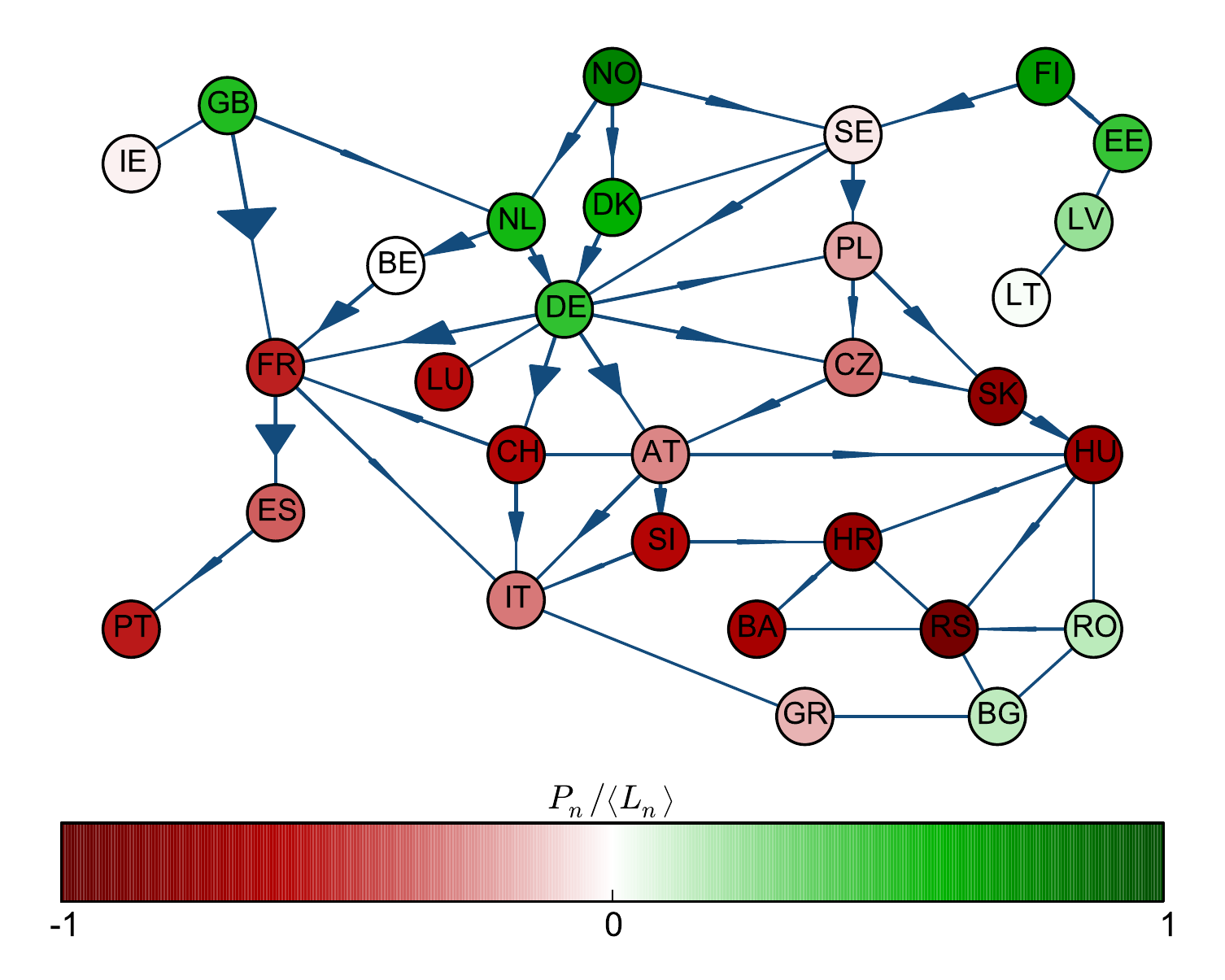}
\caption{
Injection pattern and resulting power flows on a simplified European transmission network~\cite{rod14}. A typical winter night hour has been picked from the Renewable Energy Atlas developed in \cite{hei10}.
}
\label{fig:1}     
\end{figure}

In a renewable electricity network, like the one illustrated in~Figure~\ref{fig:1}, each node~$n$ comes with a renewable power generation~$G_n^R(t)$ and a load~$L_n(t)$. Their temporal averages~$\langle G_n^R \rangle = \gamma_n \langle L_n \rangle$ define the renewable penetration~$\gamma_n$. Even for~$\gamma_n=1$ the renewable power generation at a specific time~$t$ almost certainly will not match the load. It requires a balancing infrastructure to deal with this mismatch:
\begin{equation}
\label{eq:two1}
   G_n^R(t) - L_n(t)  =  B_n(t) + P_n(t)  \; .
\end{equation}
The nodal balancing~$B_n(t)$ describes the backup power generation $G_n^B(t) = -\min\{B_n(t),0\}$ when $B_n(t)<0$, and the curtailment $C_n(t) = \max\{B_n(t),0\}$ of excess power when $B_n(t)>0$. A positive injection $P_n(t)>0$ allows the node to export power to other nodes in the network and to reduce its curtailment. In case of a negative injection $P_n(t)<0$ the node is importing power from other nodes, which reduces its own backup power generation. The sum $\sum_{n=1}^N P_n(t) = 0$ over all nodal injections has to be zero, representing flow conservation over the network with lossless power flows. $N$ represents the number of nodes in the network.

Equation~(\ref{eq:two1}) can be understood as an ``actio = reactio'' equation. The mismatch on the left-hand side drives the networked system. The right-hand side represents the response of the system. In principle, more response types can be added to the right-hand side, like storage and coupling to the heating and transportation sectors. 

There are various ways how to divide the mismatch on the left-hand side of equation~(\ref{eq:two1}) onto the balancing and the injection. A particularly simple and appealing interaction between~$B_n$ and~$P_n$ is described by synchronized balancing~\cite{rod15}:
\begin{equation}
\label{eq:two2}
   B_n(t)  =  \left( \sum_{m=1}^N \left( G_m^R(t) - L_m(t) \right) \right) \frac{\langle L_n \rangle}{\sum_k \langle L_k \rangle}  \; .
\end{equation}
It distributes the total mismatch onto each node in proportion to its average load. Together with equation~(\ref{eq:two1}) the synchronized balancing~(\ref{eq:two2}) fixes the injection pattern~$P_n(t)$. The constraint $\sum_{n=1}^N P_n(t) = 0$ is fulfilled automatically. 

The injection pattern determines the active power flow~$F_{l}$ along the links~$l=(m{\to}n)$ \cite{wood14}:
\begin{eqnarray}
\label{eq:two3}
   P_n(t)  &=&  \sum_m B_{nm}\theta_m (t)   \; , \\
\label{eq:two4}
   F_{m\to n}(t)  &=&  b_{mn} (\theta_m(t)-\theta_n(t))   \; .
\end{eqnarray}
%
Here, the DC approximation has been evoked, where link resistances can be neglected~\cite{wood14, stott09}. As in~\cite{rod15}, for convenience all reactances are set to one and the dependence of the reactances on line length is disregarded, as this leads only to slightly modified results. The susceptance matrix $B_{nm} = (\sum_k b_{nk})\delta_{nm}-b_{nm}$ then simply becomes identical to the Laplacian of the network graph, where $b_{nm}=1$ if nodes $n$ and $m$ are directly connected by a link and $b_{nm}=0$ otherwise. The variables~$\theta_n$ represent the voltage phase angles.

Figure~\ref{fig:1} illustrates the nodal injections and the resulting flows on the links in a simplified European transmission grid for one time instance. Data from a Renewable Energy Atlas~\cite{hei10} have been used for the renewable power generation $G_n^R(t) = G_n^W(t) + G_n^S(t)$, where spatio-temporal wind velocity and solar radiation fields have been converted into country-specific long-term wind and solar power generation time series with hourly resolution. Historical load data provides the demand or load $L_{n}(t)$ per country with the same temporal resolution~\cite{hei10}. For all the results shown here in this paper a fixed renewable penetration parameter~$\gamma_n=1$ and a fixed wind fraction $\langle G_n^W \rangle / \langle G_n^R \rangle = 0.8$ have been used. 

Sampling over time produces backup power distributions~$\mathcal{P}_n(G_n^B)$ for the nodes and flow distributions~$\mathcal{P}_l(F_l)$ for the links. The averages~$\langle G_n^B \rangle$ of the nodal distributions describe the required average backup energy. High quantiles of these distributions describe the required backup capacities, which secure the power supply almost surely at all times. Backup energy and capacity have been discussed for example in~\cite{rod15,hei11}. Here, we are interested in the required transmission link capacities~$\mathcal{K}_l^T$ for unconstrained power flow through the network. They follow from the high quantiles of the distributions for the absolute link flows~$f_l = |F_l|$:
\begin{equation}
\label{eq:two5}
   q  =  \int_{0}^{\mathcal{K}_l^T(q)} \mathcal{P}_l(f_l) df_l  \; .
\end{equation}
We adopt~$q=0.99$. Other definitions for the transmission link capacities have been given for example in~\cite{rod14,bec14}. The sum over all link capacities 
\begin{equation}
\label{eq:two6}
   \mathcal{K}^T   =   \sum_{l=1}^L \mathcal{K}_l^T
\end{equation}
determines the overall transmission capacity of the network. $L$ represents the number of links in the network. 

Based on the data from the Renewable Energy Atlas~\cite{hei10} the overall sum~(\ref{eq:two6}) turns out to be eight times bigger than the overall current interconnector strenghts (Net Transfer Capacities) across Europe. For some of the links this ratio becomes even larger. Together with the average nodal loads~$\langle L_n \rangle$, the specific values for the transmission link capacities have already been listed in~\cite{rod14}. Although a flow scheme different from~(\ref{eq:two2}) has been used there, the resulting transmission link capacities are qualitatively similar to the present ones. The big increase in overall transmission capacities raises the question on who has to pay for this investment. A power flow tracing based network usage measure helps to clarify this point.

%

\section{Flow tracing}
\label{sec:3}

For a grid usage measure, the link capacity $\mathcal{K}^T_l$ should be related to those nodes who are using the link~$l$ for their own power exports and imports. The fraction $\mathcal{K}^T_{ln}$ associated with node~$n$ should be proportional to its link usage. Flow tracing allows to determine this link usage by either following where the nodal exports $(P_n(t))_+ = \max\{P_n(t),0\}$ are flowing to, or by tracing back where the nodal imports $(P_n(t))_- = -\min\{P_n(t),0\}$ are coming from. Power flow tracing has been first proposed in~\cite{bia96} using the principle of proportional sharing; see also~\cite{sha02,kir97,str98}. Using the language of physics, flow tracing is related to a vector diffusion process on a directed flow graph embedded in a network. In the following we explain the details.

$F_{ln}(t)$ describes the part of the flow~$F_l(t)$ along link $l$, which has originated in the exporting node~$n$. Their ratio is denoted as $c_{ln}(t) = F_{ln}(t)/F_l(t)$ and forms one component of the colour vector $\vec{c}_l(t) = (c_{l1}(t), \ldots, c_{lN}(t))^T$ of link $l$; here the superscript $T$ stands for the transpose of the vector, and not for power transmission as it has been used for example in equation (\ref{eq:two6}). The one-norm of this vector is equal to $\| \vec{c}_l \|_1 = \sum_n c_{ln} = 1$. 

The colour vector along the directed link~$l=(m{\to}n)$, which has the same direction as the flow~$F_l(t)$, is identical to the colour vector of the upstream node~$m$, $\vec{c}_l(t) = \vec{c}_m(t)$. With other words, the upstream node copies its colour vector onto all of its outgoing flows $F_{m\to n}$ and $(P_{m})_{-}$, where $n$ are direct neighbours of $m$. Here we have used the notation $\vec{c}_{l}$ for the colour vector on link~$l$, and $\vec{c}_{n}$ for the colour vector on node~$n$. Note that both types of vectors have $N$ components denoted as $c_{lm}$ and $c_{nm}$, respectively, which correspond to the part of the (out-)flow originating in node $m$.

The colour vector $\vec{c}_{m}$ of node~$m$ is determined by merging all of its ingoing flows and their respective colour vectors:
\begin{eqnarray}
\label{eq:three3}
   \sum_{k\in\mathcal{N}_{m}^\mathrm{in}} F_{k{\to}m} \vec{c}_k + (P_m)_+ \vec{e}_m 
     =  F_{m}^\mathrm{in} \vec{c}_m  \; ,
\end{eqnarray}
where
\begin{equation}
\label{eq:three4}
   F_{m}^\mathrm{in} 
      =  \sum_{k\in\mathcal{N}_{m}^\mathrm{in}} F_{k{\to}m} + (P_m)_+ 
\end{equation}
represents the total in-flow into node~$m$. The sum goes over all direct neighbours~$k$, which produce ingoing flows for~$m$. $\vec{e}_m$ is a unit vector with components~$e_{mi}=\delta_{mi}$.

This procedure represent a colour diffusion on a directed network without loops. Equation~(\ref{eq:three3}) mixes all colour vectors flowing into a node, and the mixed colour vector is then copied onto and distributed via all of the outgoing flows. 

\begin{figure}
\includegraphics[width=0.33\textwidth]{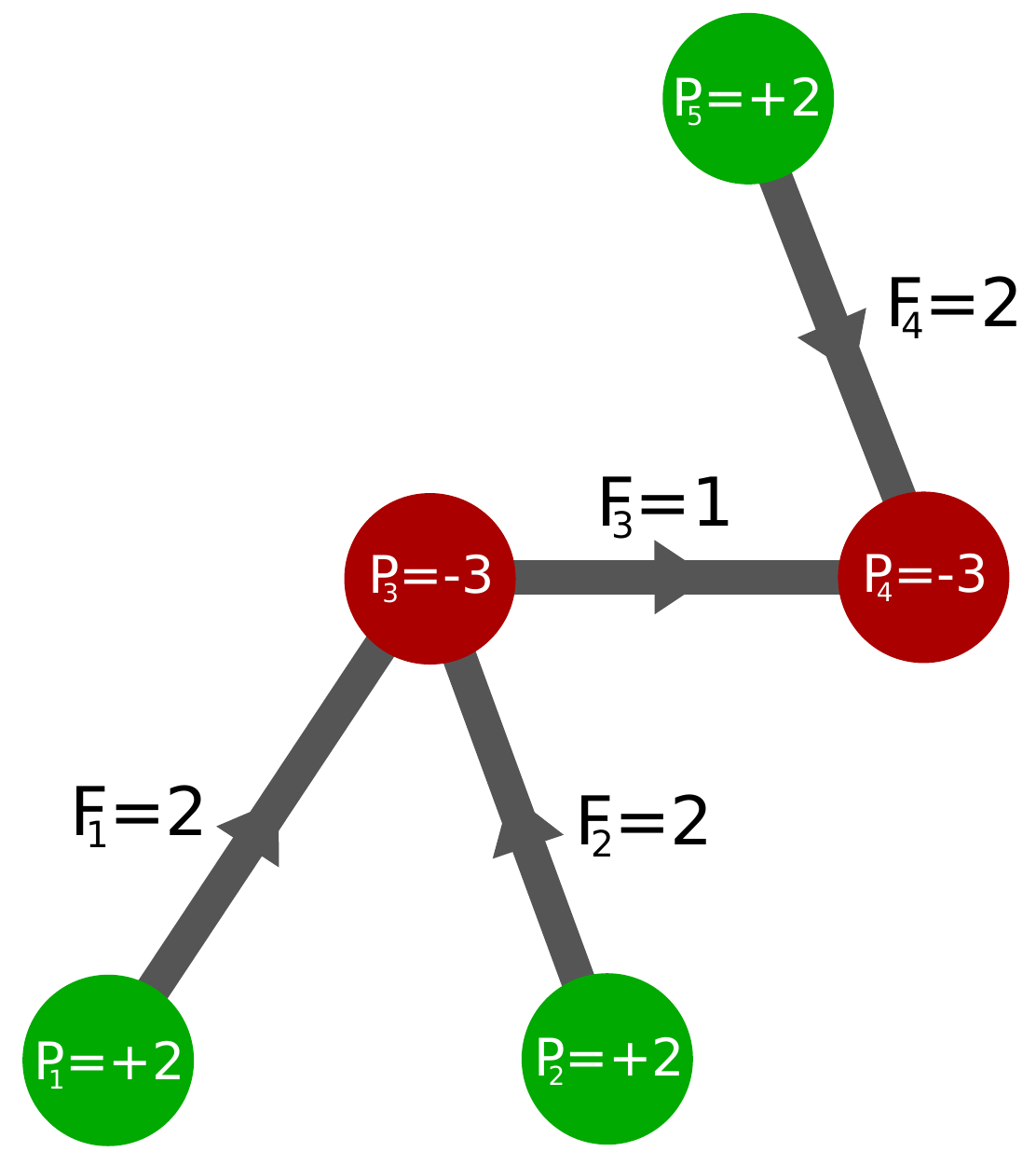}\\
\includegraphics[width=0.23\textwidth]{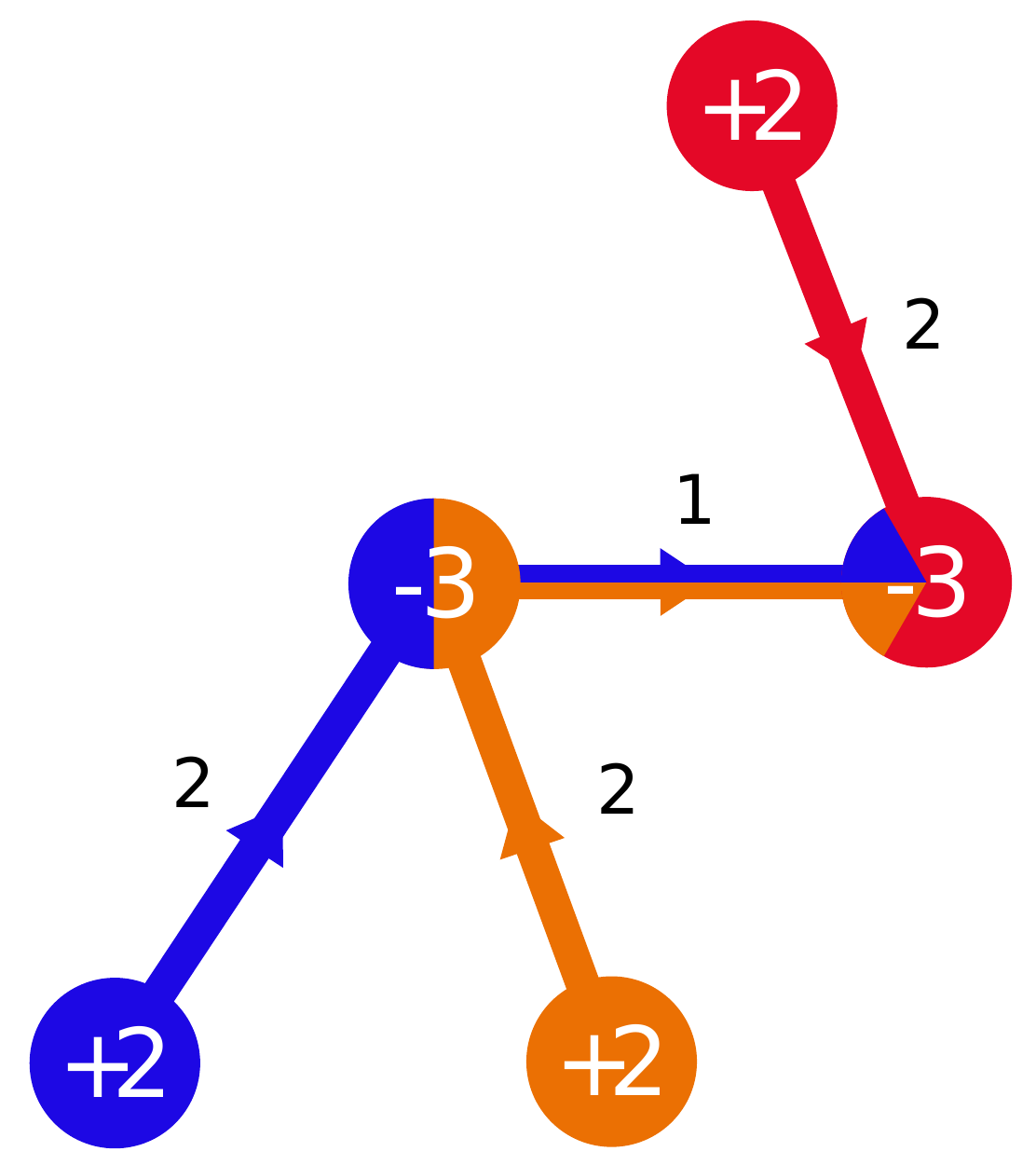}
\includegraphics[width=0.23\textwidth]{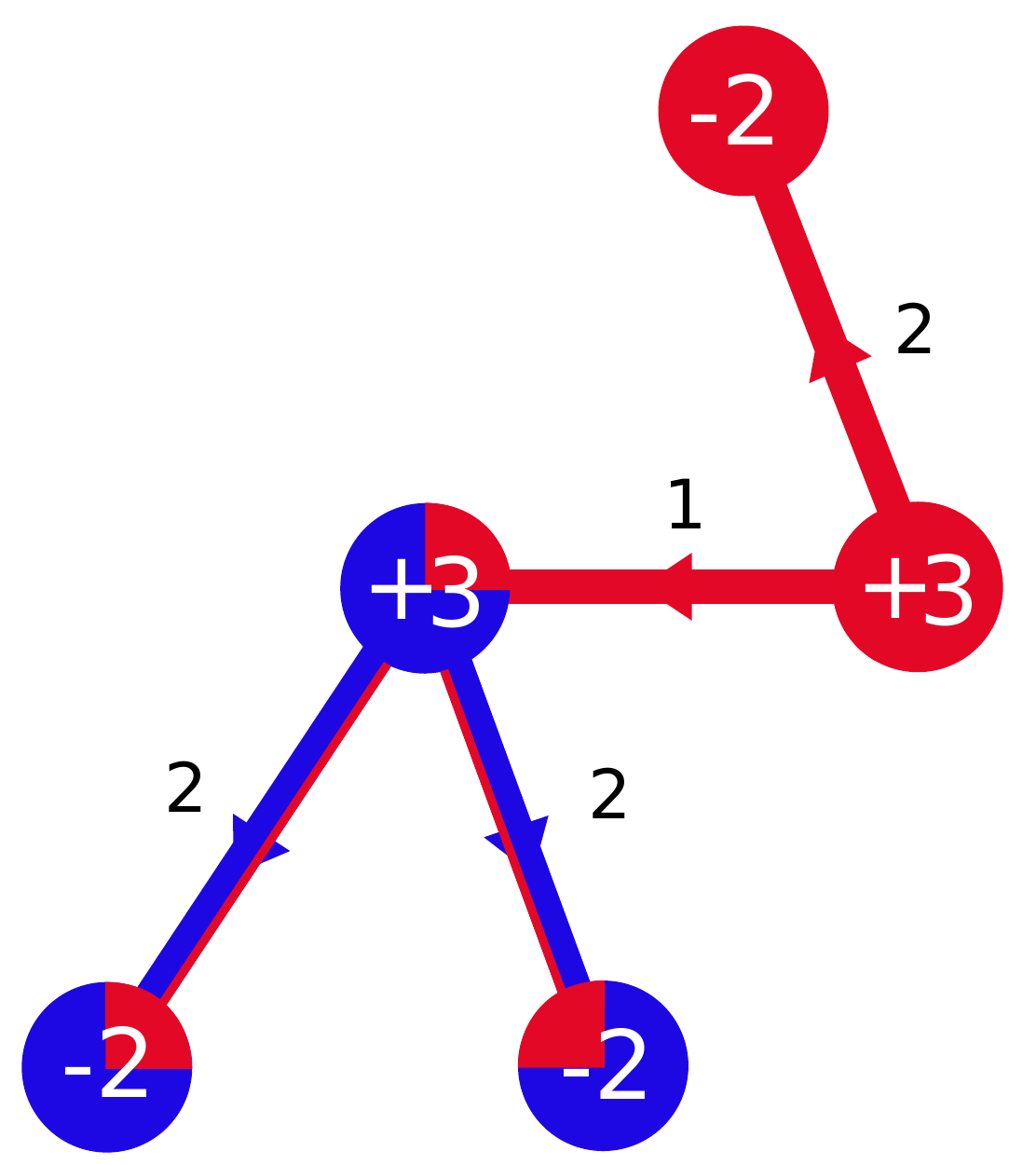}
\caption{
Illustration of a simple five-node network with
(a) an injection and a resulting flow pattern, (b) the flow tracing of exports, and (c) the flow tracing of imports.
}
\label{fig:2}      
\end{figure}

The algorithmic solution of equation~(\ref{eq:three3}) for the nodal colour vectors uses a downstream procedure. It is exemplified in Figure~\ref{fig:2}a+b. The first node $n=1$ injects $P_1=2$ units into the network. Since it has no ingoing flows on links, the only ingoing colour is the node's own injection. This results in the colour vector $\vec{c}_1 = (1,0,0,0,0)^T$, where the first component has been given the colour blue in Figure~\ref{fig:2}b. Analogously, we get the colour vectors $\vec{c}_2=(0,1,0,0,0)^T$ and $\vec{c}_5=(0,0,0,0,1)^T$ for the second and fifth node, which are like the first node only pure upstream nodes with positive injections. The colours of the second and fifth component are illustrated in yellow and red, respectively. The colour vectors of the first, second and fifth node are copied onto the first, second and fourth link, respectively. The third node has ingoing flows along the first two links and no own positive injection. This fixes the colour vector of the third node to $\vec{c}_3=(1/2,1/2,0,0,0)^T$, which is shown as the nodal 50\% blue / 50\% yellow mix in Figure~\ref{fig:2}b. This colour vector is then copied onto link~$l=3$, which carries the link-based outflow of the third node. The node~$n=4$ receives ingoing flows via the links~$l=3$ and~4, and has no positive nodal injection. The weighting of the two ingoing colour vectors with the respective flow strengths then yields the colour vector $\vec{c}_4=(1/6,1/6,0,0,2/3)^T$. 

Figure~\ref{fig:2}b illustrates the export picture of flow tracing, where link usage is caused by exporting nodes. This can of course also be turned around to the import picture of flow tracing. In this dual picture, the role of the sources and sinks in the injection pattern is exchanged, i.e.\ $P_n(t) \to -P_n(t)$. It is illustrated in Figure \ref{fig:2}c, and allows to define link usage caused by importing nodes.

%

\section{Application of flow tracing to a simplified highly renewable European electricity network}
\label{sec:4}

\begin{figure}
\includegraphics[width=0.5\textwidth]{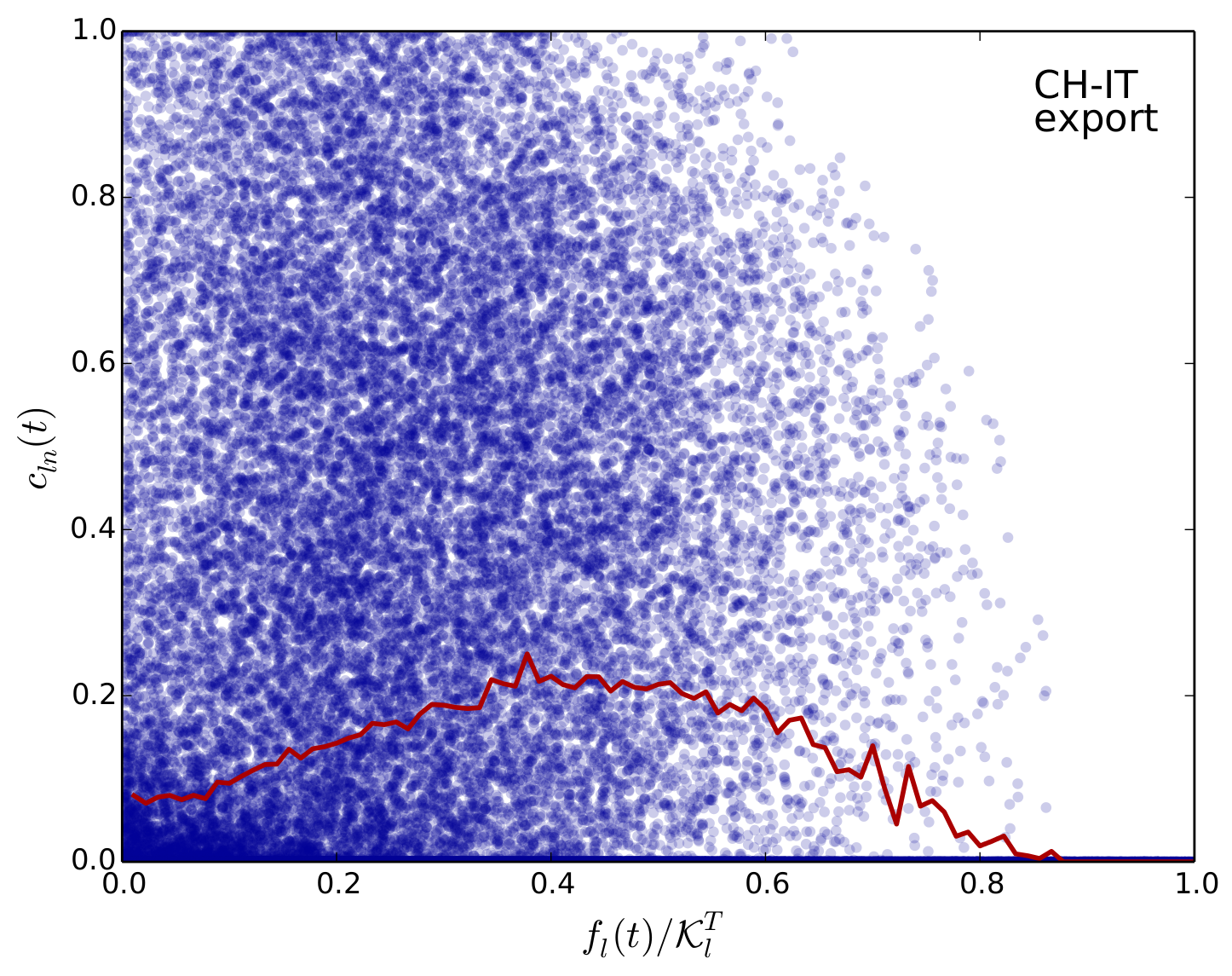}
\includegraphics[width=0.5\textwidth]{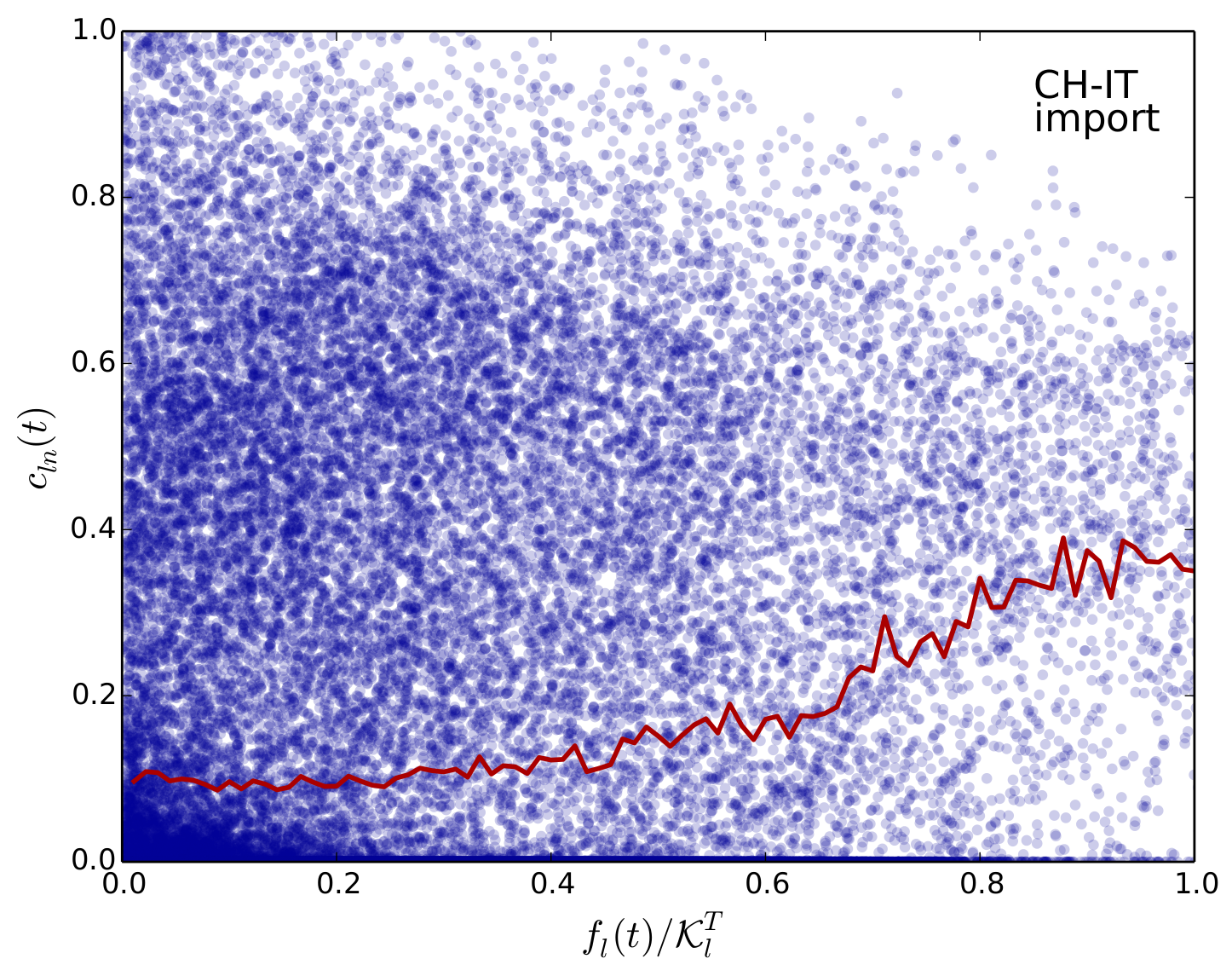}
\caption{
Colour component $c_{ln}(t)$ as a function of the power flow $f_l(t)=|F_l(t)|$: (a) based on the flow tracing of exports, and (b) based on the flow tracing of imports. The link $l=$ (CH,IT) between Switzerland and Italy, and the node $n=$ DE (Germany) have been chosen. Each dot represents one hour contained in the 8 years long Renewable Energy Atlas data from \cite{hei10}. The red curve represents the conditional average $\langle c_{ln} | f_l \rangle$. For the latter, the extreme flow events $f_l>\mathcal{K}_l^T$, which happen in $1-q=1\%$ of the time, have been mapped onto $f_l=\mathcal{K}_l^T$.
}
\label{fig:3}      
\end{figure}

The mismatch between the renewable power generation and the load on the left-hand side of equation~(\ref{eq:two1}) changes with each time step. This causes also the nodal balancing~$B_n(t)$, the nodal injection~$P_n(t)$, the power flows~$F_l(t)$ on the links, and the colour vectors~$\vec{c}_n(t)$ and $\vec{c}_l(t)$ of the nodes and links to depend on time. Figure~\ref{fig:3} shows the time-sampled values for one colour component $c_{ln}$, which have been obtained  with the 8 years long hourly Renewable Energy Atlas data from~\cite{hei10}. About half of the time the values are $c_{ln}=0$, which is because node~$n$ is not injecting (ejecting) in the export (import) picture. For the other half the values appear to scatter all over. The figur shows that in general there are various injection patterns~$P_n(t)$ which for a given sampling resolution yield the same absolute flow $f_{l}(t)$ on a link, but different colour components $c_{ln}(t)$. The conditional average $\langle c_{ln} | f_l \rangle$ over the respective ``slices'' corresponding to the same absolute link flow $f_{l}$ is also shown in Fig.~\ref{fig:3}. It shows a smooth dependence on $f_{l}$, which is different for different~$n$. Note however, that the nodal sum is always equal to $\sum_{n=1}^N \langle c_{ln} | f_l \rangle = 1$. Based on this conditional average we will now derive an expression for a fair nodal assignment of link capacity usage.

Imagine the link capacity $\mathcal{K}^T_l = \int_0^{\mathcal{K}_l^T} d\mathcal{K}$ being built by increments $d\mathcal{K}$. Every increment $d\mathcal{K} = \sum_n d\mathcal{K}_n$ is used by several nodes, and decomposes into the nodal usage contributions $d\mathcal{K}_n = w_{ln}(\mathcal{K}) d\mathcal{K}$. The fraction $\mathcal{K}_{ln}^T$ of the overall link capacity $\mathcal{K}^T_l = \sum_n \mathcal{K}_{ln}^T$ used by node~$n$ then becomes
\begin{equation}
\label{eq:four1}
   \mathcal{K}_{ln}^T
      =  \int_0^{\mathcal{K}_l^T} w_{ln}(\mathcal{K}) d\mathcal{K}
          \; .
\end{equation}
As we go from capacity~$\mathcal{K}$ to $\mathcal{K}+d\mathcal{K}$, all flows larger than~$\mathcal{K}$ make use of this additional capacity increment~$d\mathcal{K}$. Thus, the weight can be expressed as
\begin{equation}
\label{eq:four2}
   w_{ln}(\mathcal{K})
      =  \int_{\mathcal{K}}^{\mathcal{K}_l^T} \mathcal{P}_l(f_l | f_l>\mathcal{K}) \langle c_{ln} | f_l \rangle df_l
          \; ,
\end{equation}
where the conditional probability
\begin{equation}
\label{eq:four3}
   \mathcal{P}_l(f_l | f_l>\mathcal{K}) 
      =  \frac{ \mathcal{P}_l(f_l) }{ 1-\mathcal{P}^{C}_l(\mathcal{K}) }
          \; ,
\end{equation}
is proportional to the flow distribution~$\mathcal{P}_l(f_l)$, and $\mathcal{P}^{C}_l(\mathcal{K}) = \int_0^{\mathcal{K}} \mathcal{P}_l(f_l) df_l$ represents the cumulative flow distribution. Insertion of~(\ref{eq:four2}) and~(\ref{eq:four3}) into~(\ref{eq:four1}) leads to the expression 
\begin{equation}
\label{eq:four4}
   \mathcal{K}^T_{ln}
      =   \int_0^{\mathcal{K}^T_l} \frac{d\mathcal{K}}{1-\mathcal{P}^{C}_l(\mathcal{K})}
           \int_{\mathcal{K}}^{\mathcal{K}^T_l} \mathcal{P}_l(f_l) \langle c_{ln} | f_l \rangle df_l
\end{equation}
for the fraction of link capacity used by node~$n$. It is straightforward to check that this assignment fulfills $\sum_n \mathcal{K}^T_{ln} = \mathcal{K}^T_l$. 

\begin{figure*}
\resizebox{0.75\textwidth}{!}{\includegraphics{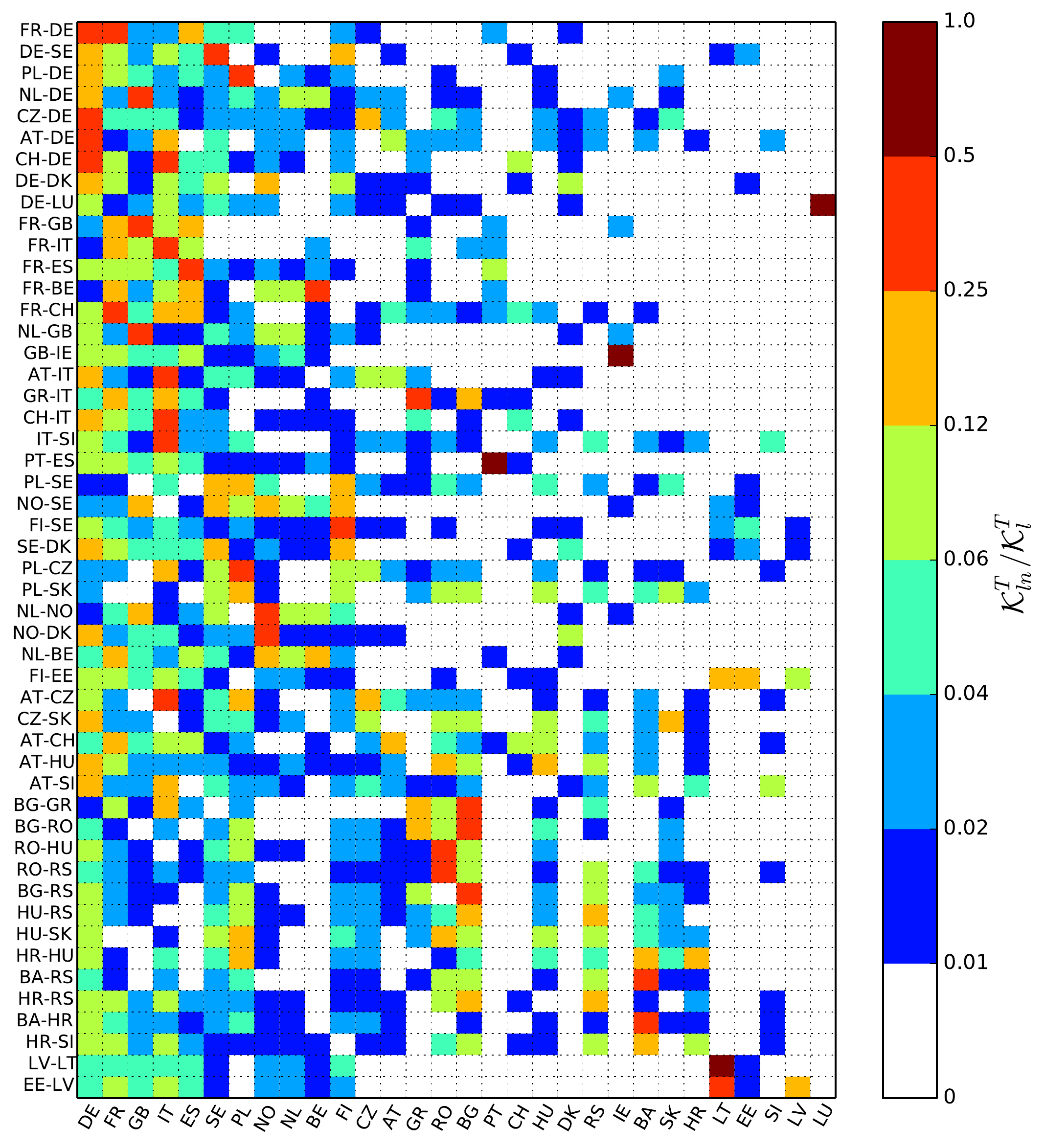}}
\caption{
$L \times N$ matrix of the relative nodal link capacities $\mathcal{K}^T_{ln}/\mathcal{K}^T_l$ resulting from a 50\%/50\% combination of the export / import pictures of flow tracing. The $L=50$ links are labeled on the left and the $N=30$ nodes are labeled on the bottom. Each row sums up to unity.
}
\label{fig:4}      
\end{figure*}

Based on the 8~years long hourly Renewable Energy Atlas data from~\cite{hei10}, the resulting relative link capacities $\mathcal{K}^T_{ln}/\mathcal{K}^T_l$ are summarized in Figure \ref{fig:4}. Each row in this figure represents a link, and each column represents a nodal country. The countries have been ordered according to their average load~$\langle L_n \rangle$, with Germany having the largest load and Luxembourg the smallest.

\begin{figure*}
\resizebox{0.49\textwidth}{!}{\includegraphics{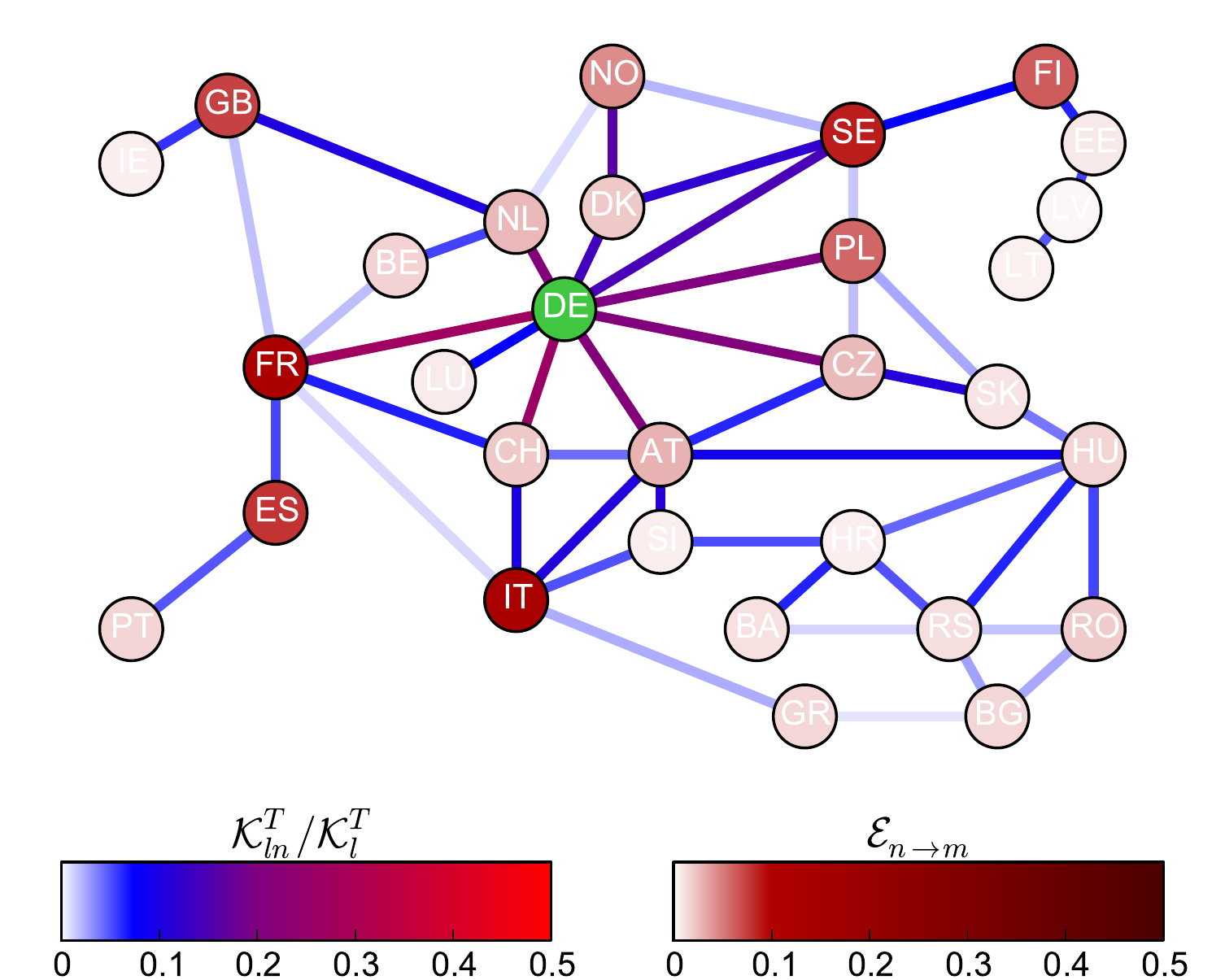}}
\resizebox{0.49\textwidth}{!}{\includegraphics{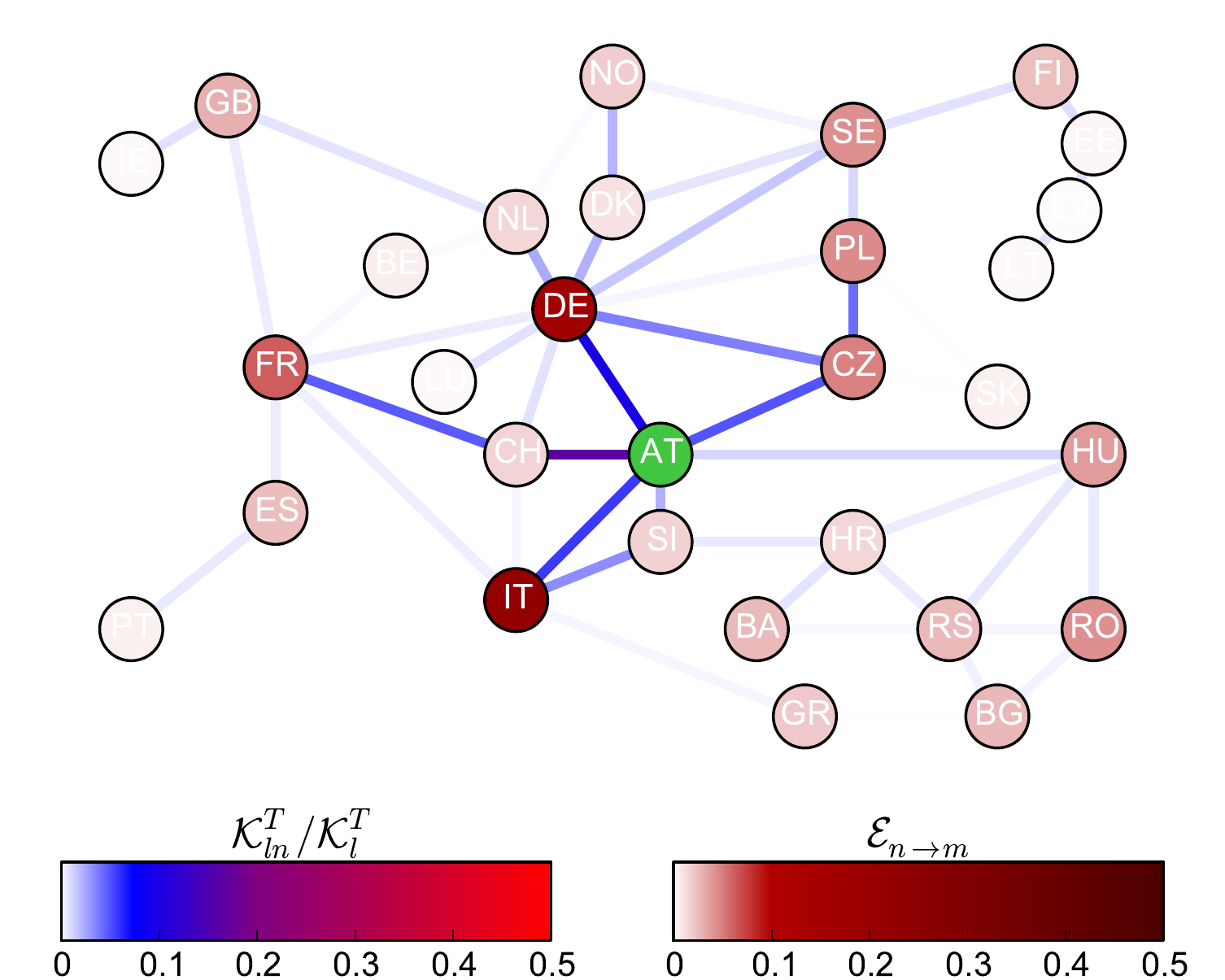}}
\caption{
Relative link capacity $\mathcal{K}^T_{ln}/\mathcal{K}^T_l$ in the export picture for the nodes DE=Germany (left) and AT=Austria (right), shown as color-coded links. Also shown is the average export transfer function (\ref{eq:three5}), shown as color-coded nodes.
}
\label{fig:5}      
\end{figure*}

Germany is using all the links contained in the network. This is also illustrated in Figure \ref{fig:5}a. The next biggest countries still use almost all of the links. The smaller a country becomes, the less number of links are used; see also Figure \ref{fig:5}b, which illustrates the case of small-medium sized Austria. Intuitively this is clear. Due to their big size, large countries are able to inject / eject large exports / imports, which then penetrate the whole network. A quantitative measure for this penetration is described by the average nodal export and import transfer functions:
\begin{eqnarray}
\label{eq:three5}
   \mathcal{E}_{n{\rightarrow}m}(t)  &=&  \frac{ \langle c_{mn} (P_m)_- \rangle }{ \langle (P_n)_+ \rangle }  \; , \\
\label{eq:three6}
   \mathcal{I}_{n{\leftarrow}m}(t)  &= & \frac{ \langle c_{nm} (P_n)_-\rangle }{ \langle (P_n)_- \rangle }  \; .
\end{eqnarray}
The former describes how much of the positive injection $(P_n)_+$ at node~$n$ is exported to a sink node~$m$. The latter expression describes how much of the negative injection $(P_n)_-$ at node~$n$ is imported from a source node~$m$. The export transfer function is also illustrated in Figure~\ref{fig:5}.  Whereas the German exports spread over the whole European network, smaller Austria mainly serves first neighbours and bigger second neighbours with its exports. The import transfer functions are not shown. They are similar to the export transfer functions. 

We now compare the flow tracing based usage measure with two simplified propositions which already have been outlined in the introduction. The first ``load-proportional'' usage~$\mathcal{M}_{n}^{(1)}$ is based on the average load of country $n$, 
\begin{equation}
\label{eq:one1}
   \mathcal{M}_n^{(1)}  =  \frac{\langle L_n \rangle}{\sum_m \langle L_m \rangle} \mathcal{K}^T  \; ,
\end{equation}
whereas the second ``link-based'' assignment $\mathcal{M}_{n}^{(2)}$ associates the capacity usage $\mathcal{K}_{l}^{T}$ of a link $l=(m \to n)$ to its incident nodes~$n$ and~$m$ only, 
\begin{equation}
\label{eq:one2}
   \mathcal{M}_n^{(2)}  =  \sum_{l(n)} \frac{\mathcal{K}^T_l}{2}  \; .
\end{equation}
Here $l(n)$ denotes all links~$l$ incident to node~$n$. Both assignments $\mathcal{M}_{n}^{(1)}$ and $\mathcal{M}_{n}^{(2)}$ add up to $\sum_n \mathcal{M}_n^{(i)} = \mathcal{K}^T$. 

\begin{figure*}
\resizebox{1.00\textwidth}{!}{\includegraphics{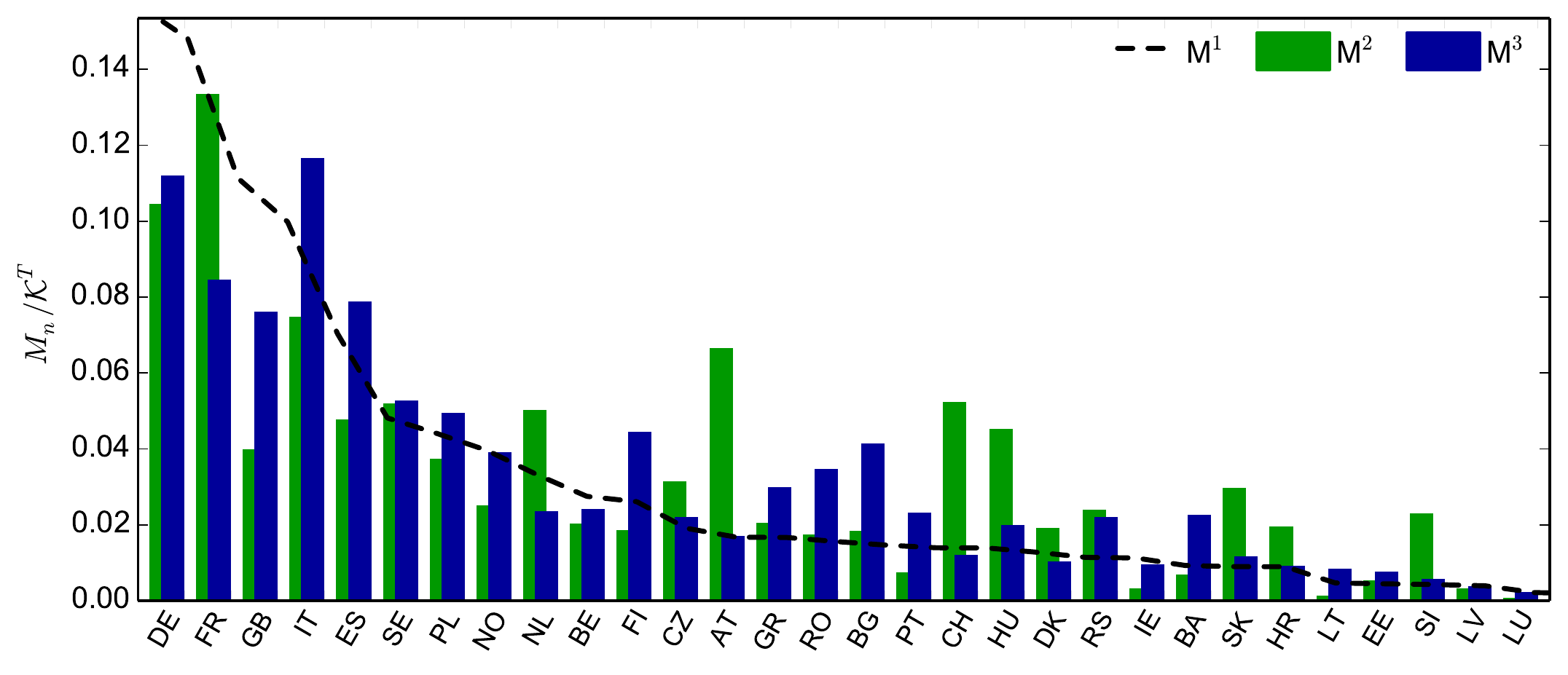}}
\caption{
Comparison of the three capacity assignments $\mathcal{M}_n^{(i)}$ from (\ref{eq:one1}), (\ref{eq:one2}) and (\ref{eq:three6}), based on a 50\%/50\% combination of export and import flow tracing.
}
\label{fig:6}     
\end{figure*}

The new flow tracing based nodal usage of the network can be expressed as the sum of nodal link capacities~(\ref{eq:four4}) over all links,
\begin{equation}
\label{eq:three6}
   \mathcal{M}_n^{(3)}  =  \sum_l \mathcal{K}^T_{ln}\; .
\end{equation}
Note that this assignment also adds up to $\sum_n \mathcal{M}_n^{(3)} = \mathcal{K}^T$. For the simplified European transmission network, $\mathcal{M}_n^{(3)}$ is illustrated in Figure~\ref{fig:6}, and compared to the two other propositions~$\mathcal{M}_n^{(1)}$ and~$\mathcal{M}_n^{(2)}$ from~(\ref{eq:one1}) and~(\ref{eq:one2}). 

For the three biggest countries Germany, France and Great Britain the new assignment~(\ref{eq:three6}) turns out to be smaller than the load-proportional assignment~(\ref{eq:one1}). For the next two countries Italy and Spain it is the opposite. This excess usage of the network becomes clearer when comparing~$\mathcal{M}_n^{(3)}$ directly to~$\mathcal{M}_n^{(2)}$. Since France is central, it has large link capacities to the other four countries, which explains the rather big green bar in Figure \ref{fig:6}. Great Britain, Spain and also Italy are peripheral. They not only use the direct links to France, but also the links beyond. This explains the reduction from~$\mathcal{M}_n^{(2)}$ to~$\mathcal{M}_n^{(3)}$ for France at the expense of an increase for Great Britain, Spain and Italy.

The Netherlands, Switzerland and Austria are central transit countries sandwiched between big countries. Consequently, their link-based assignment~$\mathcal{M}_n^{(2)}$ is exceptionally large. The usage assignment~$\mathcal{M}_n^{(3)}$ reduces these values and almost coincides with the load-proportional assignment~$\mathcal{M}_n^{(1)}$. For Hungary, Slovakia, Czech Republic, Croatia and Slovenia the situation is similar, as those serve as transits between numerous small Balkan countries on the one hand and big Germany, Italy and France on the other hand. This also explains the increased blue bars for the Balkan countries. Peripheral countries like Finland, Portugal and Ireland are in a similar situation, leading to an excess usage of the network. 

These findings highlight that the nodal link-capacity assignment~$\mathcal{M}_n^{(3)}$ based on the flow tracing network usage measure removes the unfair allocations resulting from the assignment $\mathcal{M}_n^{(2)}$ based on directly attached links. Compared to the load-proportional assignment $\mathcal{M}_n^{(1)}$, it appears that to some degree $\mathcal{M}_n^{(3)}$ reallocates capacity assignments from the big countries to medium and small-sized peripheral countries.

%

\section{Conclusion and outlook}
\label{sec:5}

Flow tracing has been developed for transmission congestion management and pricing in conventional power grids~\cite{sha02}. In its simplest form, it envokes the principle of proportional sharing~\cite{bia96}, which represents a color diffusion on a directed flow graph embedded in the network. For the first time, we have applied flow tracing to complex renewable energy networks. Those consist of nodes with fluctuating renewable power generation leading to fluctuating power flows on the links. The link capacities are determined by extreme flows and turn out to be quite large~\cite{rod14,bec14}. This poses the question of who has to pay for the link capacities. Since the flows are caused by the interaction between multiple injecting sources and ejecting sinks, the transmission capacity of a selected link is not only used by the two nodes directly attached to the link, but by more or less all nodes producing power exports and imports within the network. This topic has been and still is the subject of various debates and regulations in the European Union. For new infrastructure projects of common interest, the Agency for the Cooperation of Energy Regulators (ACER) now envisages the a~priori cross-border allocation of the cost of such projects based on a cost-benefit analysis using detailed grid models~\cite{Acer13b, ENTSOECBA}. With respect to existing grid infrastructure, transmission system operators are compensated ex-post for costs induced by transient cross-border flows. The current mechanism, which only roughly estimates the influence of imports and exports on external grids~\cite{con06}, is discussed to be replaced by a more appropriate one in the near future~\cite{Acer13a}. Flow tracing allows to derive an expression for a fair fluctuation-averaged nodal usage assignment of the link capacities. This is showcased for a simplified European transmission network with a high penetration of variable renewable power generation, where countries are treated as nodes linked by interconnectors and where data from a Renewable Energy Atlas~\cite{hei10} is used to determine the fluctuating power flows. 

This paper is intended as a first exploration of flow tracing applied to renewable energy networks. It will be interesting to extend the discussions to arbitrary renewable penetrations $\gamma_n \neq 1$. Penetrations $0 < \gamma_n < 1$ describe intermediate stages of the transition from an electricity system dominated by conventional power sources to one dominated by renewable resources, and flow tracing will allow to quantify the cost allocation of the further strengthening of the transmission grid. Furthermore it will be interesting to differentiate the flows between exporters and importers into contributions coming from wind power generation, solar power generation and conventional backup power generation. Here, flow tracing will help to allocate the future transmission grid extensions to the different forms of renewable power generation and to assign $\mathrm{CO}_2$ emission charges directly to the consumers of conventional power generation in a fair manner. 

The new flow tracing based usage assignment as presented in this paper acts as an ex post measure on a given ensemble of flow patterns in a complex renewable electricity network. Nevertheless, the method can also be used as a feedback mechanism providing incentives for future investments both in globally optimized scenarios~\cite{rod15b}, and for individual investments and operational decisions in electricity markets.

From the physics perspective, other immediate topics of interest are to relate flow tracing in larger networks with significantly more nodes and links to centrality measures of the network structure, and to study the transition from a global to a local coupling between exports and imports once transmission capacities are constrained. Also a comparison to other flow-based measures like marginal participation, which relies on the use of the Power Transfer Distribution Factor (PTDF) matrix, will be interesting, particularly with respect to the role of partial counter-flows~\cite{sha02, tb15}. Finally the relationship between power flows, diffusion processes and random walks on networks suggests to study flow tracing related concepts for these types of dynamics in abstract models, providing new insights into the fundamental properties of transport processes in complex systems.


%
\end{document}